\documentclass[11pt]{article}
\usepackage{color}
\usepackage[numbers]{natbib}
\usepackage{amstext, amssymb, amsthm, amsmath,bm,rotating,array,stmaryrd}
\usepackage{booktabs,url}
\usepackage[vlined,ruled,linesnumbered]{algorithm2e}
\usepackage[margin=1.2in]{geometry}
\usepackage{graphicx}
\usepackage{epstopdf}
\usepackage[lofdepth,lotdepth]{subfig}

\newcommand{\trans}{^{\mbox{\tiny {\sf T}}}}

\begin{document}
\title{An Online Algorithm for Nonparametric Correlations}
\author{Wei Xiao
	\vspace{0.1in}\\
	\small{\textit{SAS Institute, Cary, NC USA}}\\
	\footnotesize{wei.xiao@sas.com}}

\date{}

\baselineskip=20pt
\maketitle

\begin{abstract}
Nonparametric correlations such as Spearman's rank correlation and Kendall's tau
correlation are widely applied in scientific and engineering fields. This paper
investigates the problem of computing nonparametric correlations on the fly
for streaming data. Standard batch algorithms are generally too slow to handle
real-world big data applications. They also require too much memory because all
the data need to be stored in the memory before processing. This paper proposes
a novel online algorithm for computing nonparametric correlations. The algorithm
has $O(1)$ time complexity and $O(1)$ memory cost and is quite suitable for edge
devices, where only limited memory and processing power are available. You can
seek a balance between speed and accuracy by changing the number of cutpoints
specified in the algorithm. The online algorithm can compute the nonparametric
correlations 10 to 1,000 times faster than the corresponding batch algorithm,
and it can compute them based either on all past observations or on fixed-size
sliding windows.
\end{abstract} 

\section{Introduction}
Robust statistics and related methods are widely applied in a variety of fields
\cite{huber2011robust,rousseeuw2005robust,zaman2001econometric,xiao2016robust}. Nonparametric correlations such as
Spearman's rank (SR) correlation and Kendall's tau (KT) correlation are commonly
used robust statistics. They are often used as a replacement of the classic
Pearson correlation to measure the relationship between two random variables
when the data contain outliers or come from heavy-tailed distributions.
Applications include estimating the correlation structure of financial returns
\cite{grothe2010estimating}, comparing diets in fish \cite{fritz1974total},
and studying the relationship between summer temperature and latewood density in trees 
\cite{franceschini2013divergence}.

\begin{figure}[ht]
	\centering
	\subfloat[$Y$ is a monotonically increasing function of $X$]{
		\includegraphics[width=0.45\textwidth]{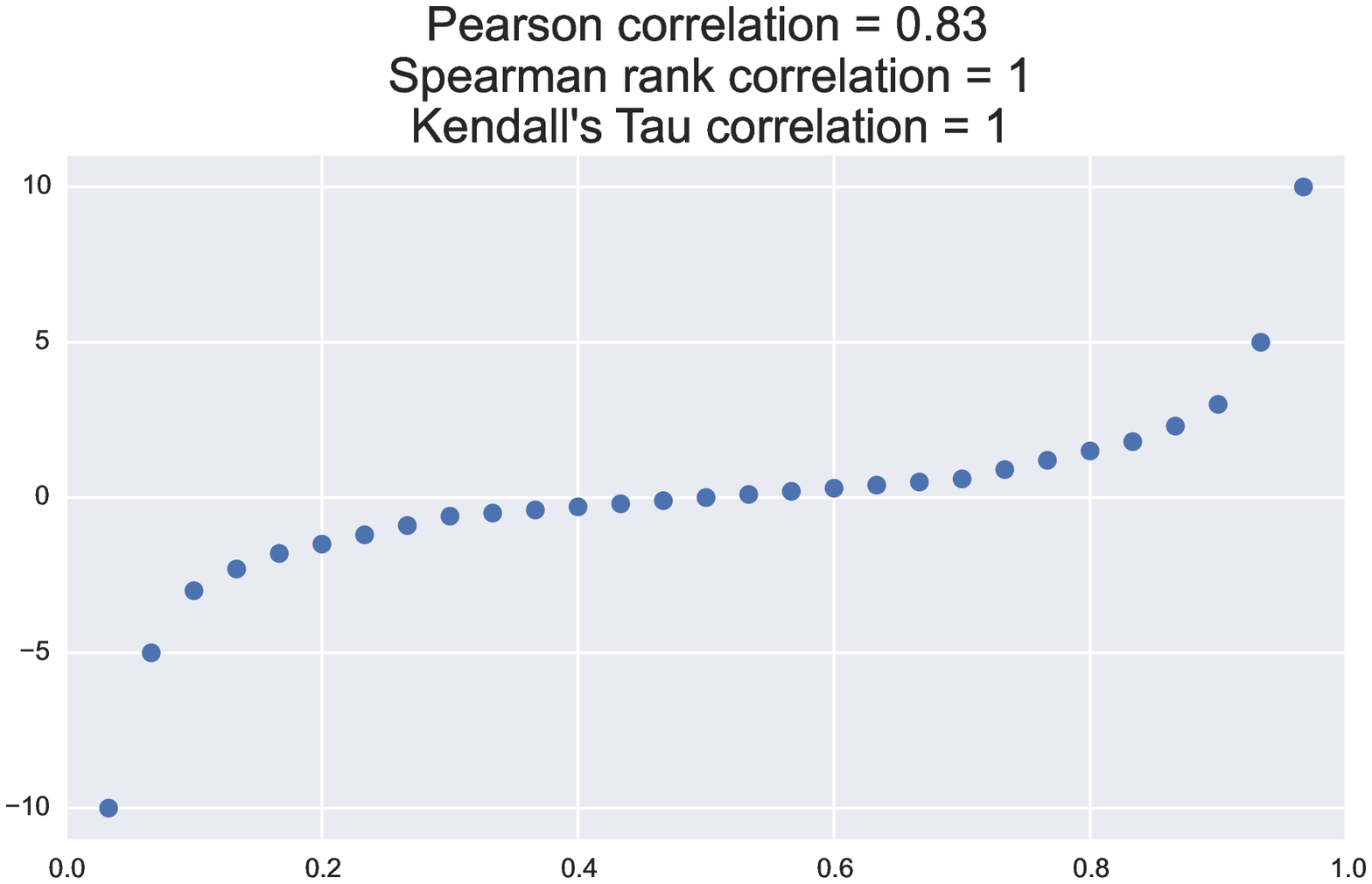}
		\label{fig:subfig1}}
	\quad
	\subfloat[Data with outliers]{
		\includegraphics[width=0.45\textwidth]{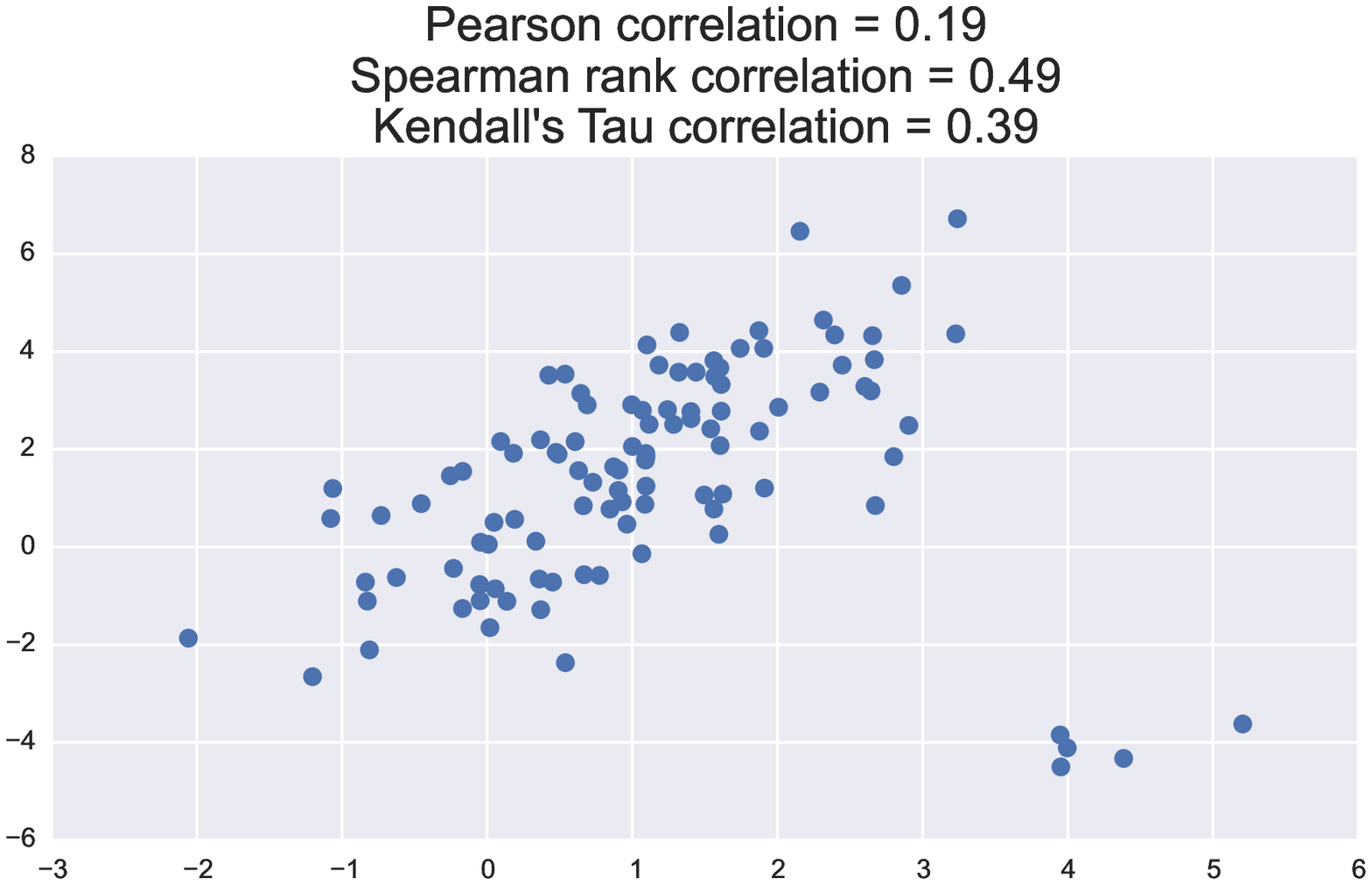}
		\label{fig:subfig2}}
	\caption{Comparison of Pearson correlation, Spearman's rank correlation, and Kendall's tau correlation on simulated data.}
	\label{fig:comparison}
\end{figure}

Nonparametric correlations have the following beneficial properties that standard
Pearson correlation does not possess. First, nonparametric correlations can work on incomplete
data (where only ordinal information of the data is available). Second, SR
and KT equal 1 when $Y$ is a monotonically increasing function of $X$. Third,
SR and KT are more robust against outliers or heavy-tailed errors. The
latter two properties are demonstrated in Figure~\ref{fig:comparison}.
Previous works have shown that the influence function of Pearson correlation
is unbounded whereas the influence functions of SR and KT are both bounded
\cite{croux2010influence,devlin1975robust}. This fact proves that Pearson correlation lacks
robustness. Furthermore, even though the Pearson correlation is the most efficient
(in teams of asymptotic variance) for a normal distribution, the efficiency
of ST and KT are both above 70\% for all values of the population correlation
coefficient \cite{croux2010influence}.

One drawback of SR and KT compared with Pearson correlation is that they require
more computational time. The computation of SR and KT requires sorting (finding
rank) of the $X$ and $Y$ sequences, which is a very time-consuming step when the
sample size $n$ is large. The minimum time complexities for batch algorithms
for SR and KT are $O(n\log n)$ \cite{knight1966computer}, whereas the time
complexity for batch algorithm of Pearson correlation is $O(n)$.

In practice you sometimes would want to analyze correlation between variables
in dynamic environments, where the data are streaming in. These environments
include network monitoring, sensor networks, and financial analysis. A good
algorithm should make it easy to incorporate new data and process the input
sequence in a serial fashion. Such algorithms are called online algorithms in
this paper. Online algorithms have interesting applications in various fields
\cite{crammer2003ultraconservative,oza2005online,gama2010knowledge,xiao2017online}.
A standard online algorithm exists to compute Pearson correlation by using
the idea of sufficient statistics \cite{gama2010knowledge}. The time complexity of 
this algorithm is $O(1)$, and its memory cost is also $O(1)$.
However, because Pearson correlation
is not robust against outliers, it is not the desirable method for some
applications, for example, suppose you collect data from a huge sensor network
of a complex system and you want to analyze the correlation in order to detect highly
correlated sensor pairs. Outliers in sensor readings might occur because of noise,
different temperature conditions, or failures of sensors or communication. Pearson
correlation would not be robust enough to handle such an analysis.

This type of analysis demonstrates the need for an online algorithm for nonparametric
correlations (such as SR and KT). However, the way the Pearson correlation is computed 
cannot be directly carried over to SR and KT.
It cannot be carried over for SR because new data can change the ranks of all historical
observations that were used to compute the correlation. For KT on
the other hand, new data need to be compared with all historical data in
the computation of the correlation. In order to compute SR and KT
with streaming data exactly, it is necessary to keep all previous history in memory, which
is impossible because the data streams can be unbounded in length.

This paper proposes an efficient online algorithm for SR and KT. The time
complexity of this algorithm is $O(1)$, and its memory cost is also $O(1)$.
Although the algorithm only approximately computes SR and KT, this paper
shows through extensive simulation studies and real applications that the
approximation is good enough for most cases. To the limit of the authors'
knowledge, the algorithm developed in this work is the first online algorithm
for nonparametric correlation.

\section{Online Algorithms for Nonparametric Correlations}
Let $\{(x_i, y_i), i \geq 1\}$ denote the streaming inputs of two time series $x$ and $y$. At time $t$, the Pearson
correlation ($r_{P}$), Spearman's rank correlation ($r_{SR}$), and Kendall's tau correlation ($r_{KT}$) computed based on all previous observations are defined as:
\begin{align*}
r_{P}\triangleq& \cfrac{\sum_{i=1}^{t}(x_i - \bar{x})(y_i - \bar{y})}{\sqrt{\sum_{i=1}^{t}(x_i-\bar{x})^2\sum_{i=1}^{t}(y_i-\bar{y})^2}}\\
r_{SR}\triangleq& \cfrac{\sum_{i=1}^{t}(u_i - \bar{u})(v_i - \bar{v})}{\sqrt{\sum_{i=1}^{t}(u_i-\bar{u})^2\sum_{i=1}^{t}(v_i-\bar{v})^2}}\\
r_{KT}\triangleq& \cfrac{P-Q}{\sqrt{(P+Q+T)*(P+Q+U)}}
\end{align*}
where $u_i$ is the rank of $x_i$, $v_i$ is the rank of $y_i$, $P$ is the number of concordant pairs, $Q$ is the number
of discordant pairs, $T$ is the number of ties only in $x$, and $U$ is the number of ties only in $y$.

\begin{figure}[ht]
	\centering
	\includegraphics[width=0.8\textwidth]{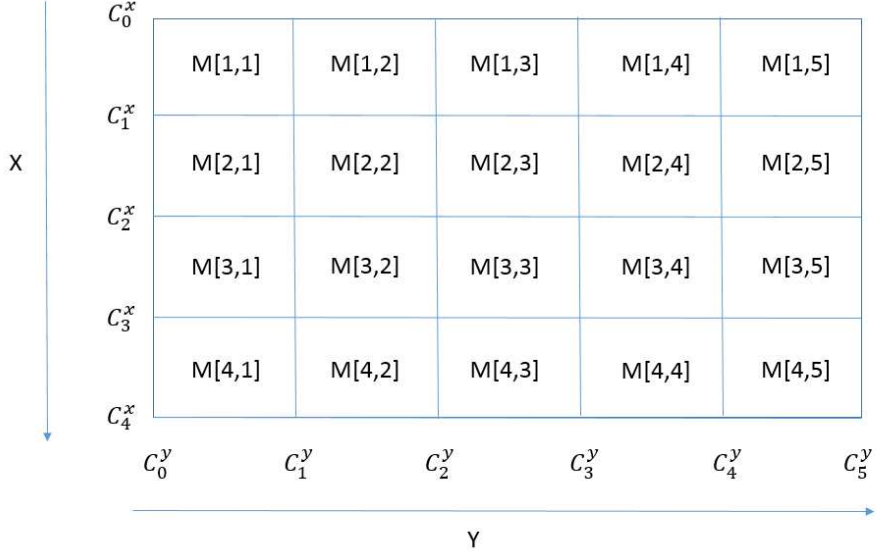}
	\caption{Example of count matrix $M$ with three cutpoints of $x$ and four cutpoints of $y$.}
	\label{plot:count_matrix}
\end{figure}

The main idea in designing the online algorithms for nonparametric
correlations is to coarsen the bivariate distribution of ($X$,$Y$).
The coarsened joint distribution can represented by using a small matrix.
Assume that both $X$ and $Y$ are continuous random variables. You provide
$m_1^*$ and $m_2^*$ distinct cutpoints in ascending order for $X$ and
$Y$ respectively, where both $m_1^*$ and $m_2^*$ are nonnegative integers. The
cutpoints for $X$ are denoted as $(c^x_1$,\ldots,$c^x_{m_1^*})\trans$,
where $c^x_1<c^x_2<\ldots<c^x_{m_1^*}$. Similarly, the cutpoints for $Y$
are represented as $(c^y_1$,\ldots,$c^y_{m_2^*})\trans$. Two default
cutpoints are added for $X$: $c^x_0$ and $c^x_{m_1^*+1}$, where $c^x_0=-\infty$ and
$c^x_{m_1^*+1}=\infty$. Cutpoints $\{c^x_i,\,i=0\ldots,m_1^*+1\}$
discretize $X$ into $m_1$ ranges, where $m_1=m_1^*+1$. The same is done for $Y$,
and cutpoints $\{c^y_i,\,i=0\ldots,m_2^*+1\}$ discretize $Y$ into $m_2$ ranges,
where $m_2=m_2^*+1$. The $m_1 \times m_2$ count matrix $M$ is then constructed,
where $M[i,j]$ stores the number of observations that falls into the range
$[c^x_{i-1}, c^x_{i})\times[c^y_{j-1}, c^y_{j})$. An example of an $M$ matrix
is shown in Figure~\ref{plot:count_matrix}, where three cutpoints are chosen for
$X$ and four cutpoints are chosen for $Y$. Using the count matrix $M$ has two
advantages: first, instead of entire $(x_i, y_i)$ series (which
maybe unbounded in length) being stored,the information is stored in a matrix of
fixed size.
Second, when $(x_i,y_i)$ are discretized and stored in $M$, they are naturally
sorted, and fast algorithms exist to quickly compute $r_{SR}$ and $r_{KT}$
from $M$. This paper proves that the time complexity of the algorithms is
$O(m_1m_2)$ for both Spearman's rank correlation and Kendall's tau correlation.
Because both $m_1$ and $m_2$ are fixed integers, the algorithms for
both Spearman's rank correlation and Kendall's tau have $O(1)$ time complexity
and $O(1)$ memory cost. This makes the implementation of these algorithms
quite attractive on edge devices, where only limited memory and processing power
are available. In practice cutpoints for $X$ and $Y$ need to be chosen. One good
choice for cutpoints are the equally spaced quantiles of the random variable.
For example, to choose 9 cutpoints for $X$, we can use the sample quantiles of
$X$ that correspond to the probabilities $0.1, 0.2, \ldots, 0.9$.

The preceding discussion assumes that both $X$ and $Y$ are continuous. When $X$ and $Y$ are discrete or ordinal,
cutpoints can be selected so that each pair of consecutive cutpoints contains only one level of the random variable. When
both $X$ and $Y$ are discrete or ordinal no information is lost by using $M$ to approximate the bivariate distribution
of ($X$, $Y$), and the algorithm's result is the exact nonparametric correlation between $X$ and $Y$.

The general online algorithm for SR and KT is given in
Algorithm~\ref{algo:online_npcorr}. To expedite the computation, not
only is matrix $M$ tracked, but also its row sum $n_{\mathrm{row}} =
(n_{\mathrm{row}}[1],\ldots, n_{\mathrm{row}}[m_1])\trans$, its column sum
$n_{\mathrm{col}} = (n_{\mathrm{col}}[1],\ldots, n_{\mathrm{col}}[m_2])\trans$,
and its total sum $n=\sum_{i=1}^{m_1}\sum_{j=1}^{m_2}M[i,j]$. The algorithm
is also designed to compute nonparametric correlation and return the
result every $n_{\mathrm{gap}}$ new observations. The $n_{\mathrm{gap}}$
is a user-specified parameter, where $n_{\mathrm{gap}}\geq 1$. When the
observation index $t$ mod $n_{\mathrm{gap}}$ is not equal to 0, only $M$,
$n_{\mathrm{row}}$, $n_{\mathrm{col}}$, and $n$ need to be updated (Step 3--6),
and the nonparametric correlation does not need to be computed (Step 8) in the
iteration. When $t$ mod $n_{\mathrm{gap}}$ is equal to 0, it is necessary both to update 
$M$, $n_{\mathrm{row}}$, $n_{\mathrm{col}}$ and $n$, and to compute the
nonparametric correlation. Unlike the step that computes the nonparametric correlation,
the updating steps can be done very efficiently with time complexity
$O(\max(\log(m_1),\log(m_2)))$.

\begin{algorithm}[H]
	\DontPrintSemicolon
	Input: $\{(x_t, y_t)\}_{t=1}^{T}$ (streaming observations), $(c^x_1$,\ldots,$c^x_{m_1})\trans$ (cutpoints for $x$),
$(c^y_1$,\ldots,$c^y_{m_2})\trans$ (cutpoints for $y$), $n_{\mathrm{gap}}$.\;
	\For{t=1 to $T$}{
		Reveal the sample $(x_t, y_t)$.\;
		Compare $x_t$ with $(c^x_1$,\ldots,$c^x_{m_1})\trans$ and find the row index $i$ of $M$ that corresponds to the
observation via a binary search.\;
		Compare $y_t$ with $(c^y_1$,\ldots,$c^y_{m_2})\trans$ and find the column index $j$ of $M$ that corresponds to
the observation via a binary search.\;
		$M[i,j]\leftarrow M[i,j] + 1$, $n_{\mathrm{row}}[i]\leftarrow n_{\mathrm{row}}[i]+1$, $n_{\mathrm{col}}[j]\leftarrow n_{\mathrm{col}}[j] + 1$, $n\leftarrow n + 1$\;
		\If{$t$ mod $n_{\mathrm{gap}} = 0$ }{
			Compute the nonparametric correlation based on $M$, $n_{\mathrm{col}}$, $n_{\mathrm{row}}$, $n$, and save the result to $r[t]$.
		}
	}	
	\Return{$(r[n_{\mathrm{gap}}],r[2n_{\mathrm{gap}}],\ldots)\trans.$}
	\caption{General online algorithm for nonparametric correlation}
	\label{algo:online_npcorr}
\end{algorithm}

Step 8 in Algorithm~\ref{algo:online_npcorr} is described in detail for SR and
KT respectively in Algorithms~\ref{algo:online_SR} and \ref{algo:online_KT},
where the nonparametric correlations are quickly computed based on matrix
$M$, $n_{\mathrm{row}}$, $n_{\mathrm{col}},$ and $n$. It is easy to verify that the
time complexity of both algorithms is $O(m_1m_2)$ in linear proportion to the
number of cells in matrix $M$.

\begin{algorithm}[H]
	\DontPrintSemicolon
	Input: $M$, $n_{\mathrm{row}}$, $n_{\mathrm{col}}$, $n$, $m_1$, and $m_2$\;
	\# iteratively compute the rank that corresponds to each row of $M$\;
	$r = 0$\;
	\For{k=1 to $m_1$}{
		\If{$n_{\mathrm{row}}[k] = 0$}{
			$r_{\mathrm{row}}[k] \leftarrow r$
		}
		\Else{
			$r_{\mathrm{row}}[k] \leftarrow [(r+1) + (r+n_{\mathrm{row}}[k])]/2$\;
			$r\leftarrow r + n_{\mathrm{row}}[k]$
		}
	}
	\# iteratively compute the rank that corresponds to each column of $M$.\;
	$r = 0$.\;
	\For{k=1 to $m_2$}{
		\If{$n_{\mathrm{col}}[k] = 0$}{
			$r_{\mathrm{col}}[k] \leftarrow r.$
		}
		\Else{
			$r_{\mathrm{col}}[k] \leftarrow [(r+1) + (r+n_{\mathrm{col}}[k])]/2$\;
			$r\leftarrow r + n_{\mathrm{col}}[k].$
		}
	}
	$r^{*}_{\mathrm{row}}\leftarrow r_{\mathrm{row}} - (n+1)/2$; $r^{*}_{\mathrm{col}}\leftarrow r_{\mathrm{col}} -
(n+1)/2.$\;
	$r^{*}_{\mathrm{row}}\leftarrow
r^{*}_{\mathrm{row}}/\sqrt{\sum_{i=1}^{m_1}n_{\mathrm{row}}[i]r^{*}_{\mathrm{row}}[i]^2}$;
$r^{*}_{\mathrm{col}}\leftarrow
r^{*}_{\mathrm{col}}/\sqrt{\sum_{i=1}^{m_2}n_{\mathrm{col}}[i]r^{*}_{\mathrm{col}}[i]^2}.$\;
	$corr \leftarrow (r^{*}_{\mathrm{row}})\trans M r^{*}_{\mathrm{col}}.$\;	
	\Return{$corr.$}
	\caption{Compute Spearman's rank correlation based on matrix $M$}
	\label{algo:online_SR}
\end{algorithm}

\begin{algorithm}[H]
	\DontPrintSemicolon
	Input: $M$, $n_{\mathrm{row}}$, $n_{\mathrm{col}}$, $n$, $m_1$, and $m_2$.\;
	Initialize $N$ as an $m_1$ by $m_2$ zero matrix.\;
	Compute $P$ (the number of concordant pairs):\;
	\For{i=2 to $m_1$}{
		\For{j=2 to $m_2$}{
			\If{j=2}{
				$N[i,j] \leftarrow M[i-1,j-1]$
			}
			\Else{
				$N[i,j] \leftarrow N[i,j-1] + M[i-1,j-1]$
			}
		}
	}
	\For{i=2 to $m_1$}{
		$N[i,:]\leftarrow N[i,:] + N[i-1,:]$
	}
	$P = \sum_{i=1}^{m_1}\sum_{j=1}^{m_2}M[i,j]*N[i,j]$\;
	Compute $T$ (the number of ties only in x): $T = [\sum_{i=1}^{m_1}(n_{\mathrm{row}}[i]^2 - \sum_{j=1}^{m_2}M[i,j]^2)]/2$\;
	Compute $U$ (the number of ties only in y): $T = [\sum_{j=1}^{m_2}(n_{\mathrm{col}}[j]^2 - \sum_{i=1}^{m_1}M[i,j]^2)]/2$\;
	Compute $B$ (the number of ties in both x and y): $B=\sum_{i=1}^{m_1}\sum_{j=1}^{m_2}M[i,j]*(M[i,j] - 1)/2$\;
	Compute $Q$ (the number of discordant pairs): $Q = (n+1)*n/2 - P- T - U - B$\;
	$corr \leftarrow (P-Q)/\sqrt{(P+Q+T)*(P+Q+U)}$\;	
	\Return{$corr.$}
	\caption{Compute Kendall's tau correlation based on matrix $M$}
	\label{algo:online_KT}
\end{algorithm}

The parameters $(m_1, m_2)$ control a tradeoff between the accuracy and efficiency of the online algorithms. The more
cutpoints that are chosen for $x$ and $y$ (larger $m_1$, $m_2$), the more accurate the approximation of the bivariate
distribution of $(x,y)$ with $M$ and a more accurate result is generally achieved. However, increasing $m_1$ and $m_2$ also
decreases the speed of the online algorithm. Based on the extensive numerical studies of the next section, a rule-of-thumb choice of $(m_1,m_2)$ is $m_1=m_2=30$ for SR and $m_1=m_2=100$ for KT. 

In practice, the proposed online algorithms for nonparametric correlations usually work well for the following reasons:
First, when $(m_1,m_2)$ increase to infinity, the result of online algorithms converges to the true value. Second, for a reasonably large $(m_1,m_2)$, each cell of matrix $M$ represents only a very local area of the $(X, Y)$ distribution. Positive errors and negative errors can cancel each other out when summed together.

Algorithms ~\ref{algo:online_SR} and \ref{algo:online_KT} compute SR and KT over
all past data. Frequently, you are interested in computing the statistics only
over the recent past. Specifically, you might like to compute SR and KT with
a sliding window of a fixed size. Algorithm~\ref{algo:online_npcorr} can be
easily modified to deal with such cases. The only change that is needed is
to add some steps after step 6; in the added steps $M$, is updated by first
finding the row index and the column index that correspond to the observation
$(x_{t-n_{\mathrm{win}}}, y_{t-n_{\mathrm{win}}})$ and then decreasing the
corresponding cell of $M$ by 1. Here $n_{\mathrm{win}}$ represents the size of
the sliding windows. The details are left to the readers.

\section{Simulation Studies}

In the following simulation studies, equal numbers of cutpoints are always chosen for both $X$ and $Y$, and all cutpoints
are chosen as equally spaced quantiles of a standard normal distribution. We implement the batch and online nonparametric correlation algorithms in python. Users can download the package from https://github.com/wxiao0421/onlineNPCORR.git. 
 
\subsection{Simulation Study with Nonparametric Correlations Computed over All Past Observations}
This simulation evaluates the online algorithms for SR and KT (computed over all past observations) by comparing them
with the corresponding batch algorithms. The $x_i$, $i=1,\ldots,T$ are generated from an independently and indentically
distributed (iid) normal distribution $N(0,1)$. Let $y_i=(z_i+\sigma x_i)/\sqrt{\sigma^2+1}$, where $z_i$,
$i=1,\ldots,T$ are iid $N(0,1)$ random variables, which are independent of $x_i$, $i=1,\ldots,T$. It is easy to verify
that both $\{x_i\}_{i=1}^{T}$ and $\{y_i\}_{i=1}^{T}$ are iid $N(0,1)$ with a Pearson correlation coefficient of $\sigma/\sqrt{\sigma^2+1}$.

\begin{figure}[ht]
	\centering
	\subfloat[Run times with different numbers of cutpoints ($n_{\mathrm{gap}}=1$)]{
		\includegraphics[width=0.8\textwidth]{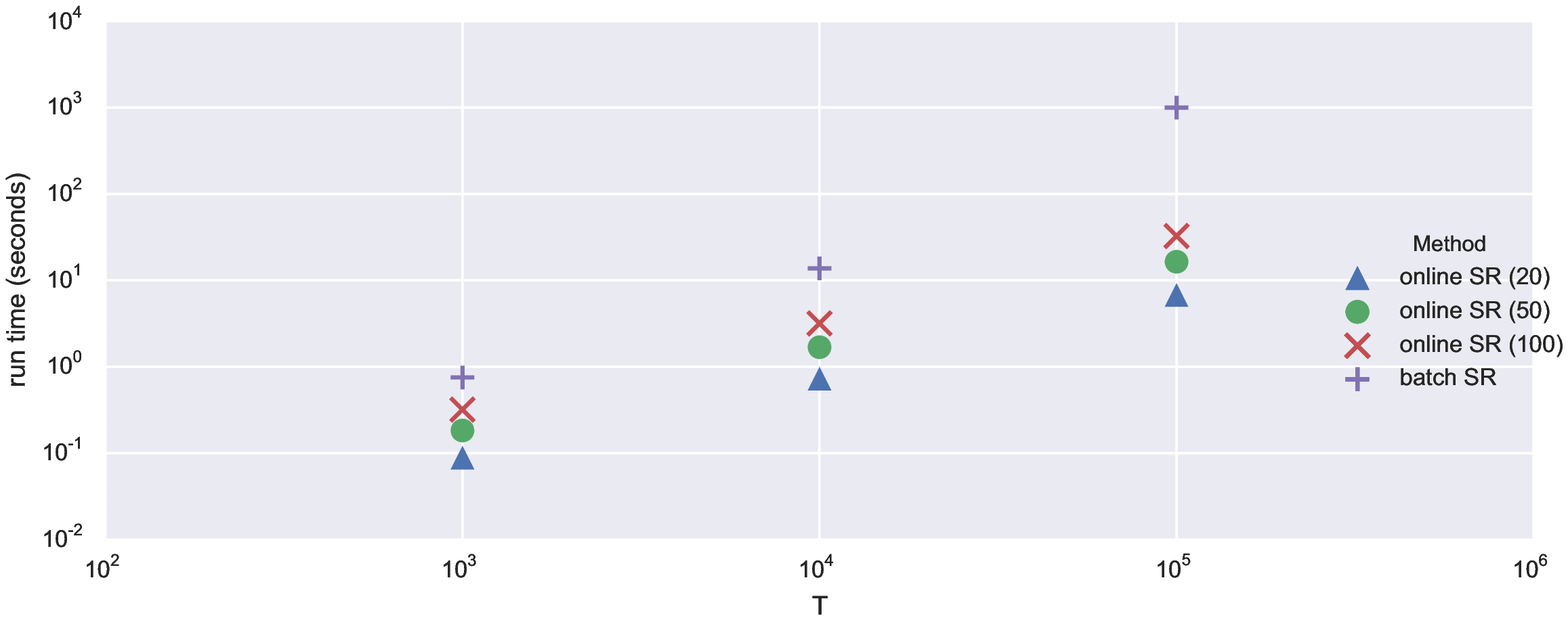}
		\label{fig:run_times_cut_points_SR_sim1}}
	\\
	\subfloat[$L_1$ errors ($n_{\mathrm{gap}}=1$)]{
		\includegraphics[width=0.8\textwidth]{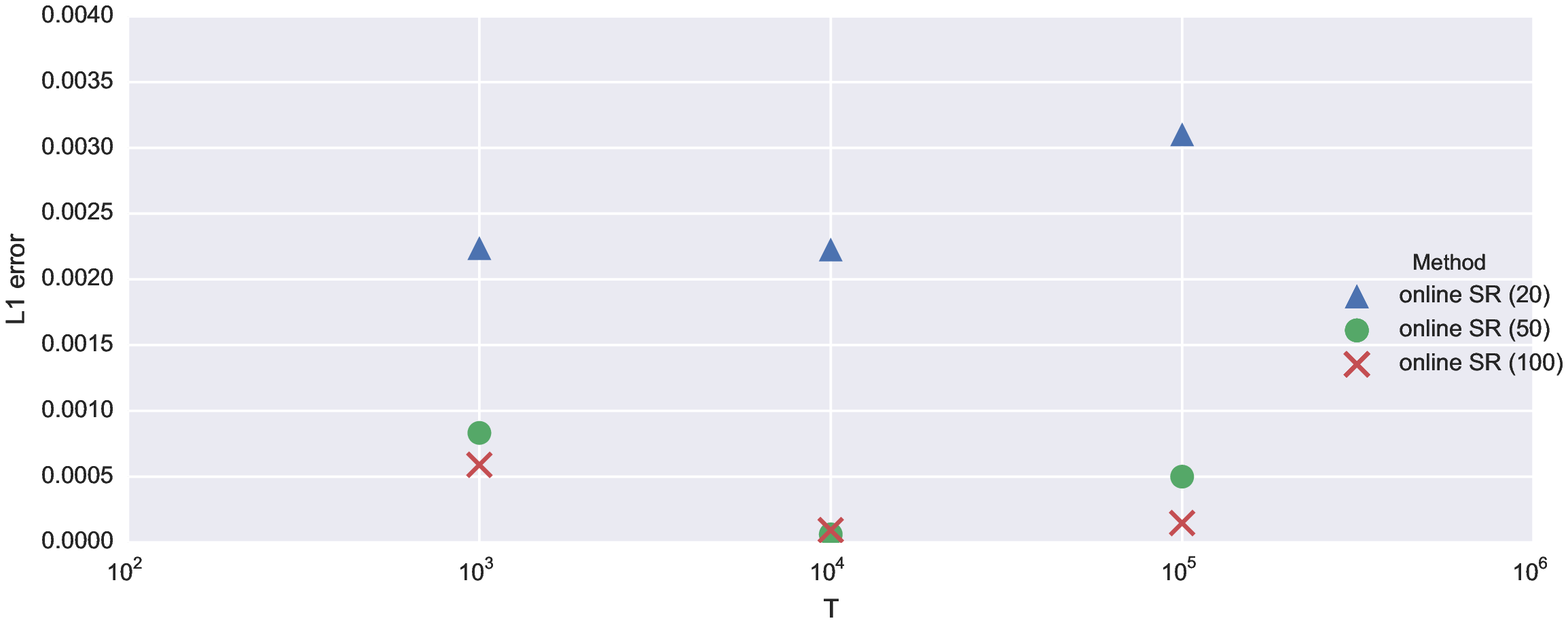}
		\label{fig:error_SR_sim1}}
	\\
	\subfloat[Run times with different $n_{\mathrm{gap}}$ (online SR (50))]{
		\includegraphics[width=0.8\textwidth]{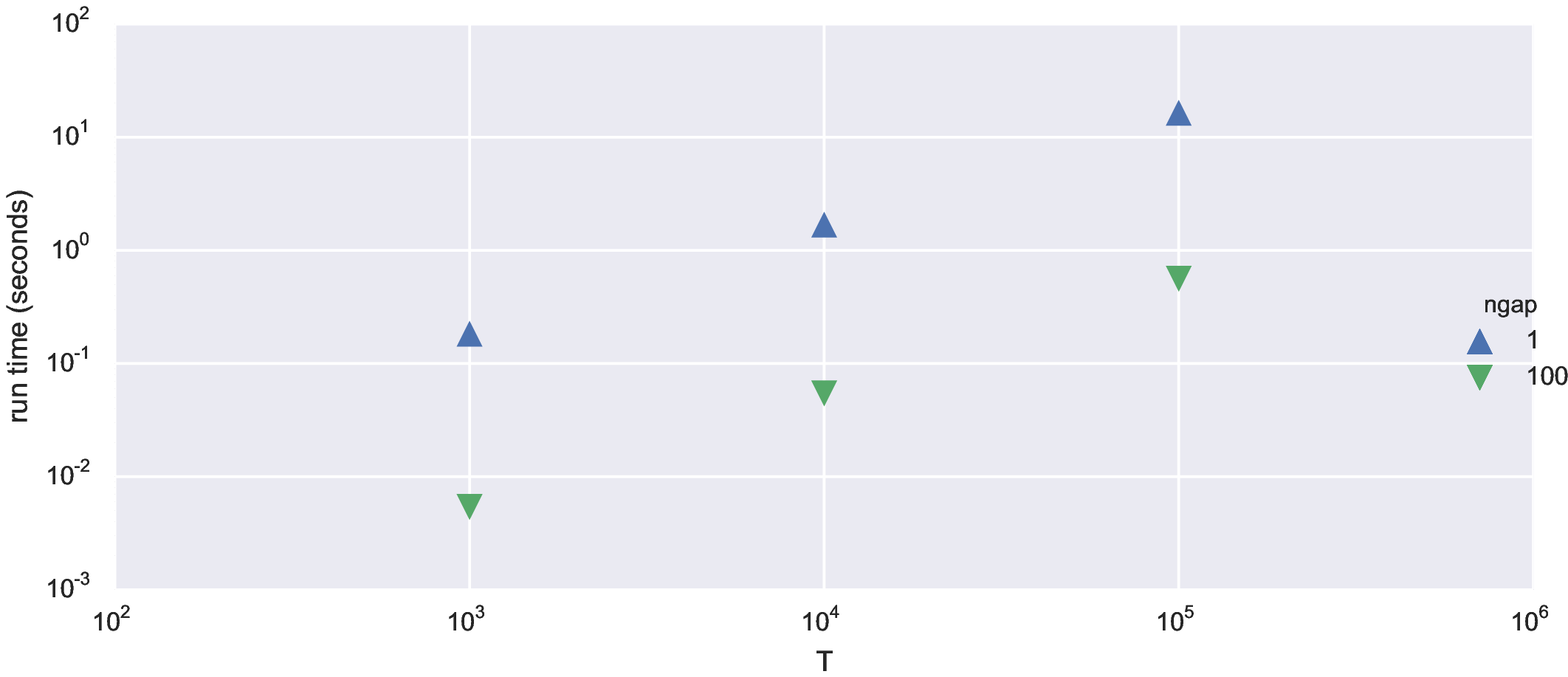}
		\label{fig:run_times_ngaps_SR_sim1}}
	\caption{Comparison of online algorithm and batch algorithm for Spearman's rank correlation, where ``online SR (K)''
represents the online algorithm with 50 cutpoints for both $X$ and $Y$ variables. All cutpoints of $X$ and $Y$ are
chosen as equally spaced quantiles of a standard normal distribution.}
	\label{fig:sim1_SR}
\end{figure}

The result for SR is shown in Figure~\ref{fig:sim1_SR}. All numbers are
averaged over 10 replications. All results in subplots (a) and (b) are based
on $n_{\mathrm{gap}}=1$. Subplot (a) compares the run times of the
batch algorithm and the online algorithms with different numbers of cutpoints.
It is clear that for the online algorithms, increasing the number of cutpoints
decreases the speed of the algorithms. Furthermore, as the number of observations
($T$) increases, the differences between the run times for the batch algorithm and
the online algorithm also increase dramatically. The SR algorithm with
20 cutpoints (online SR (20)) takes less than 10 seconds to run a $T=10^5$
case, whereas the batch algorithm takes more than 1,000 seconds. When $T$ increases to
$10^6$, the batch algorithm becomes too slow to handle such cases (time complexity
$O(T^2\log T)$), whereas it is easy to both numerically and theoretically prove
that the run time of the online algorithm is proportional to $T$. Subplot
(b) compares the L1 error of the estimated Spearman's rank correlation from
the online algorithm with different numbers of cutpoints (computed at $T$). The
L1 error does not seem to increase with $T$, and it generally decreases with the
number of cutpoints. For all cases, the L1 error is kept below 0.004. Subplot
(c) compares the run times of the online algorithm (with 50 cutpoints)
for $n_{\mathrm{gap}}=1$ and $100$. The increase in speed is 30-fold
for $n_{\mathrm{gap}}=100$ for all $T$. Last, the batch algorithm is
implemented very efficiently in C (using the Python Scipy package) whereas the
online algorithm is Purely python code. So if the online algorithm is also
implemented in C, you would likely see another 10 to 100-fold speed increase for the
online algorithm.
     
\begin{figure}[ht]
	\centering
	\subfloat[Run times with different numbers of cutpoints ($n_{\mathrm{gap}}=1$)]{
		\includegraphics[width=0.8\textwidth]{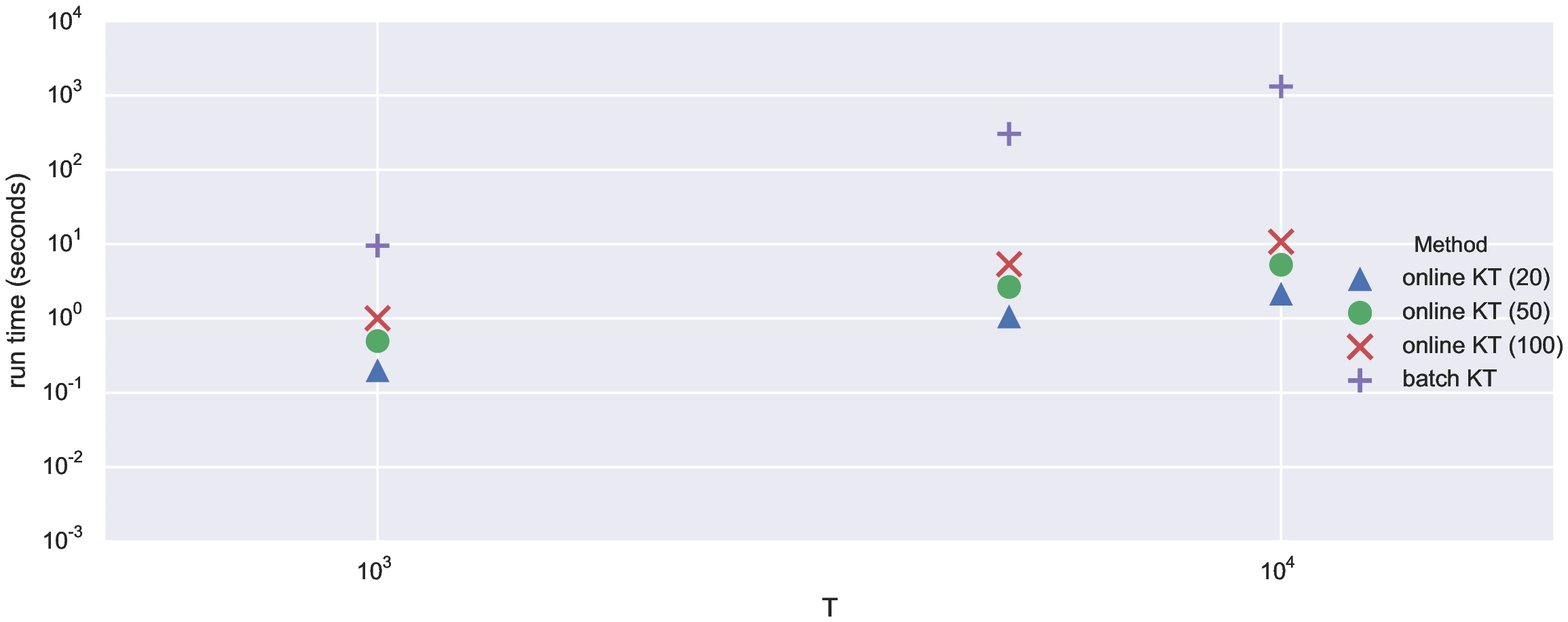}
		\label{fig:run_times_cut_points_KT_sim1}}
	\\
	\subfloat[$L_1$ errors ($n_{\mathrm{gap}}=1$)]{
		\includegraphics[width=0.8\textwidth]{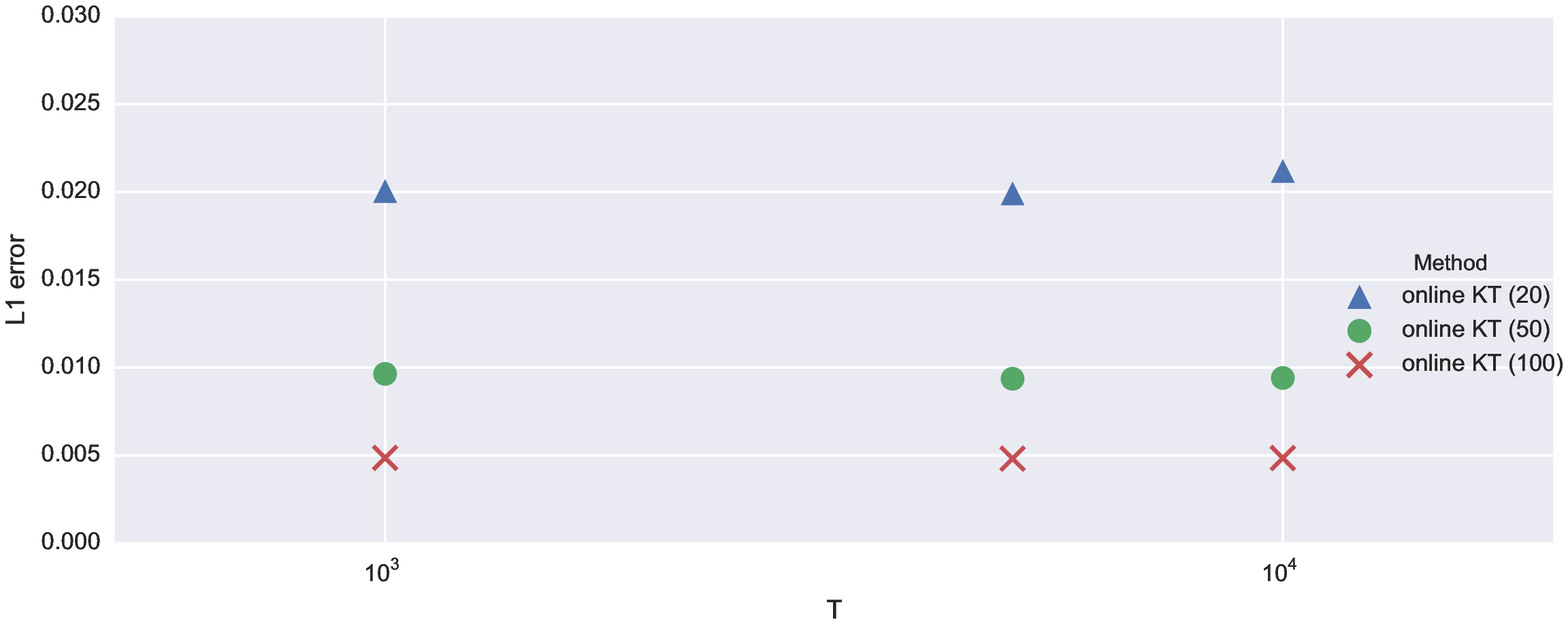}
		\label{fig:error_KT_sim1}}
	\\
	\subfloat[Run times with different $n_{\mathrm{gap}}$ (online KT (50))]{
		\includegraphics[width=0.8\textwidth]{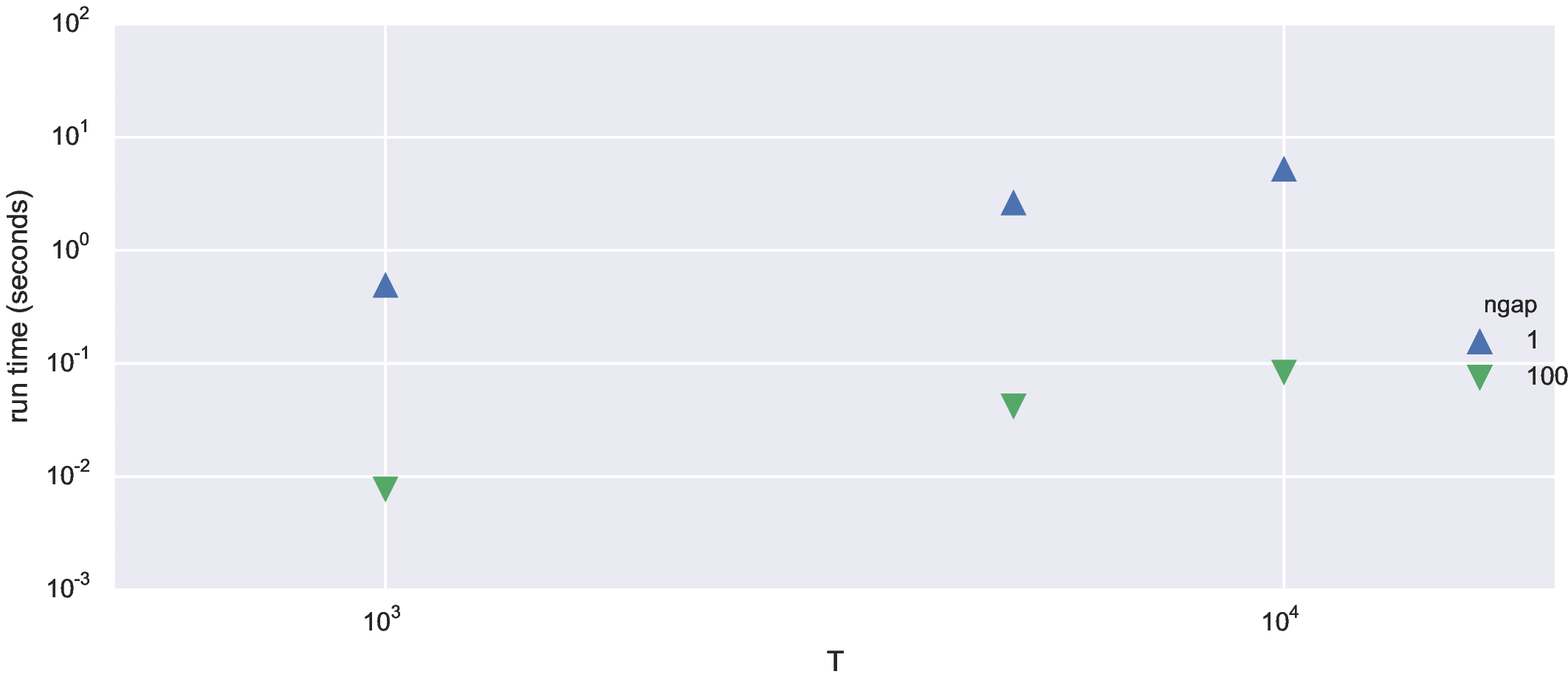}
		\label{fig:run_times_ngaps_KT_sim1}}
	\caption{Comparison of online algorithm and batch algorithm for Kendall's tau correlation, where ``online KT (K)''
represents online algorithm with 50 cutpoints for both $X$ and $Y$ variables. All cutpoints of $X$ and $Y$ are chosen as
equally spaced quantiles of a standard normal distribution.}
	\label{fig:sim1_KT}
\end{figure}

The result for KT is shown in Figure~\ref{fig:sim1_KT}. Subplot (a)
compares the run times of the batch algorithm and the online algorithm
with different numbers of cutpoints. The observed pattern is similar to that of
SR, where the run times increase with the number of cutpoints in the online
algorithm. Furthermore, as the number of observations ($T$) increases, the differences 
between the run times of the batch algorithm and online algorithm also increase
dramatically. The KT online algorithm with 100 cutpoints (online KT (100)) takes
approximately 10 seconds to run a $T=10^4$ case, whereas the batch algorithm takes
more than 1,000 seconds. When $T$ increases to $10^5$, the batch algorithm becomes
too slow to handle such cases, whereas the online algorithm can still finish
the computation in a very short period of time. Subplot (b) compares the
L1 error of the estimated Kendall's tau correlation of the online algorithm with
different numbers of cutpoints (computed at $T$). The L1 error does not seem to
change much with $T$, and it decreases with the number of cutpoints. For all cases
where the number of cutpoints is larger than 50, the L1 error is below
0.01. Subplot (c) compares the run times of the online algorithm under
$n_{\mathrm{gap}}=1$ and $100$ (with 50 cutpoints). You see increases of approximately 60
to 70 times for $n_{\mathrm{gap}}=100$ and all $T$.

\subsection{Simulation Study with Nonparametric Correlations Computed Based on Sliding Windows} 
This simulation study compares the batch and online algorithms for SR
and KT based on sliding windows. Generate $x_i$, $i=1,\ldots,T$
from an iid normal distribution $N(0,1)$. Let $y_i=(z_i+\sigma(i)
x_i)/\sqrt{\sigma(i)^2+1}$, where $z_i$, $i=1,\ldots,T$ are iid $N(0,1)$
random variables, which are independent with $x_i$, $i=1,\ldots,T$.
Then $T=100,000$ and $n_{\mathrm{win}}=10,000$ for SR, and $T=10,000$ and
$n_{\mathrm{win}}=1,000$ for KT. We choose $\sigma(i)=5[(i-m)/m]^2$, where
$m=50,000$ for SR and $m=5,000$ for KT.

\begin{figure}[ht]
	\centering
	\subfloat[Spearman's rank correlation ($n_{\mathrm{gap}}=1$)]{
		\includegraphics[width=0.8\textwidth]{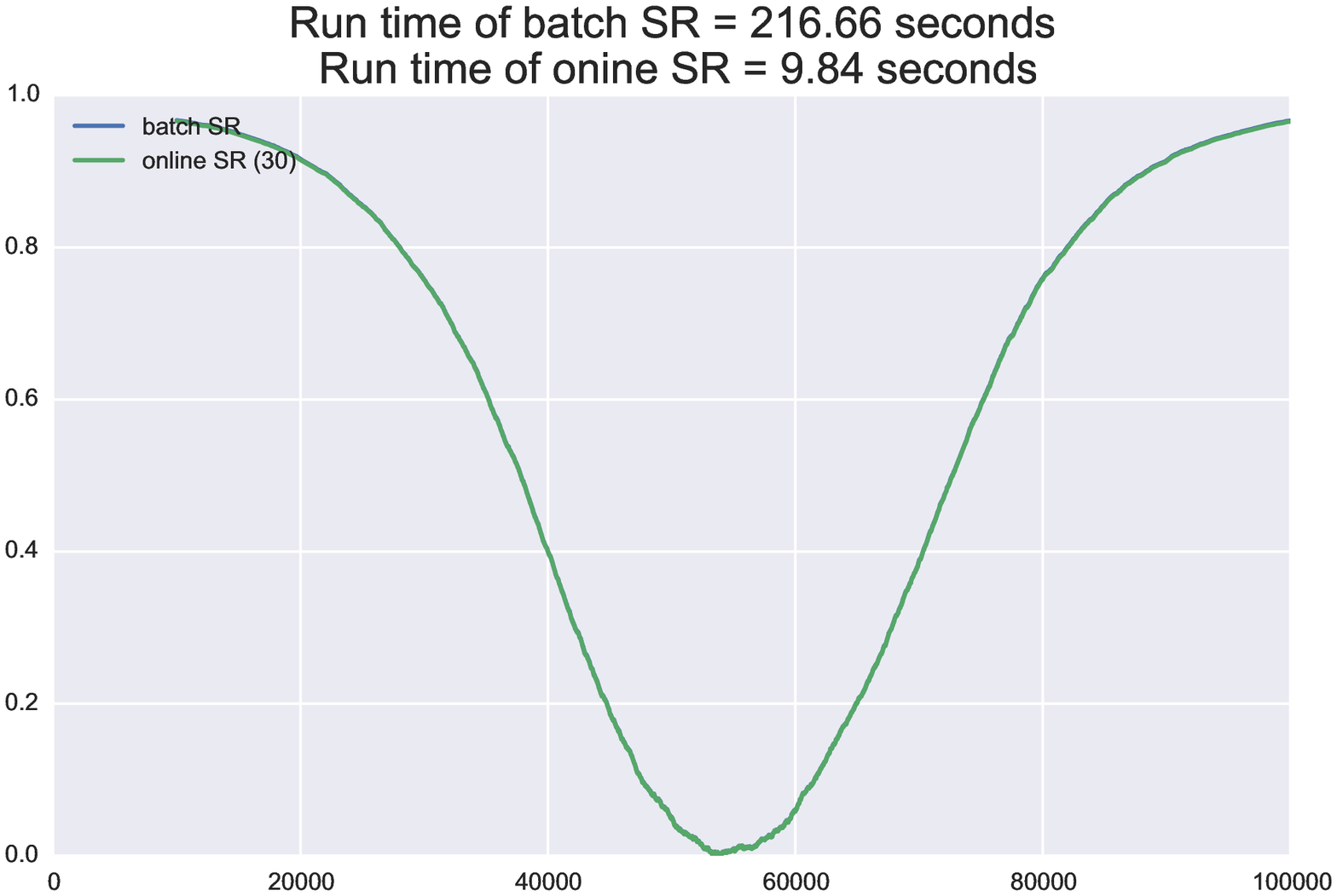}
		\label{fig:sim2_SR}}
	
	\subfloat[Kendall's tau correlation ($n_{\mathrm{gap}}=1$)]{
		\includegraphics[width=0.8\textwidth]{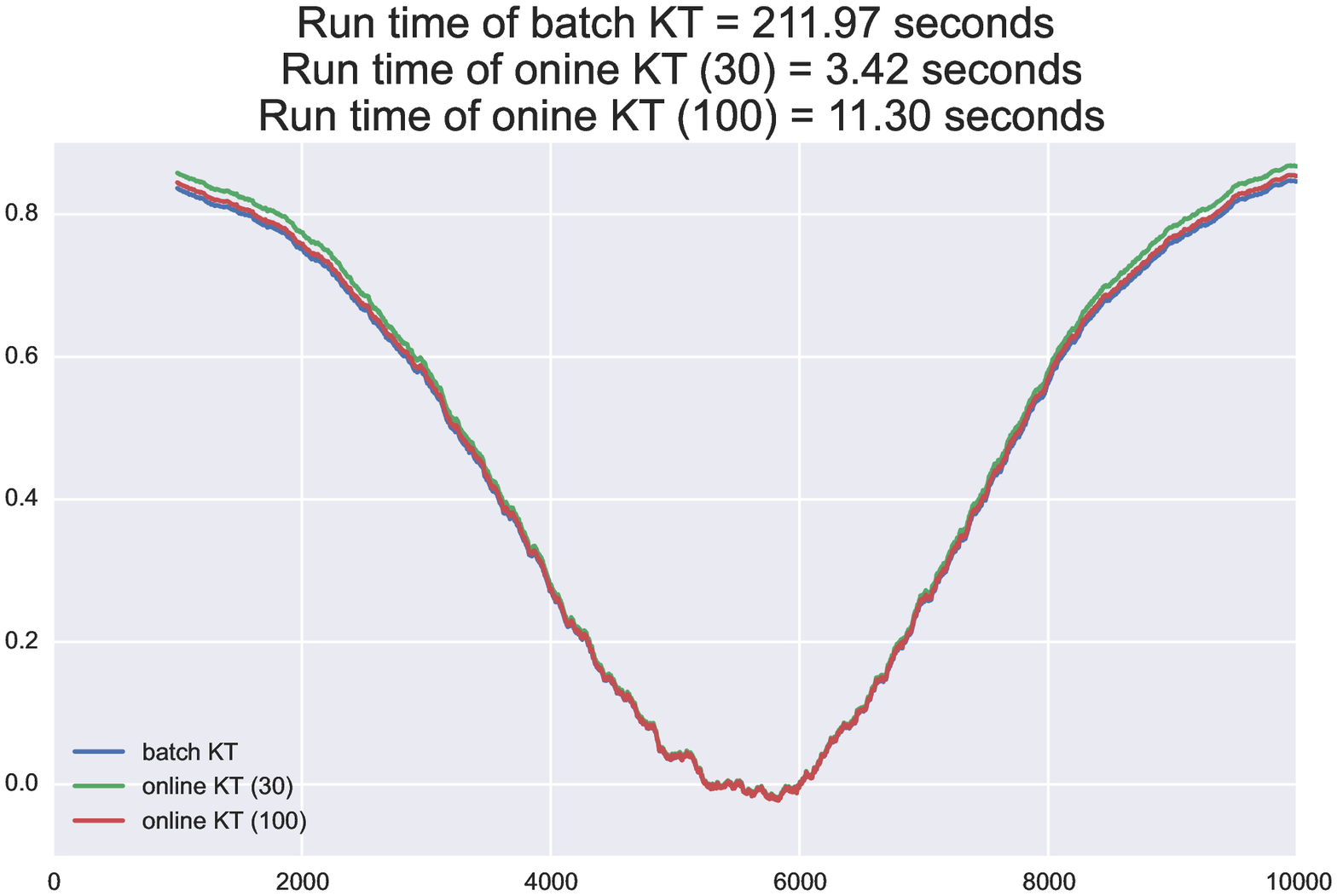}
		\label{fig:sim2_KT}}
	\caption{Comparison of online and batch algorithms for Spearman's rank correlation and Kendall's tau correlation
based on sliding windows where ``online SR (K)'' and ``online KT (K)'' represents the online algorithm for SR and KT with 50 cutpoints (for both $X$ and $Y$ variables). All cutpoints of $X$ and $Y$ are chosen as equally spaced quantiles of standard normal distribution.}
	\label{fig:sim2}
\end{figure}

The results are shown in Figure~\ref{fig:sim2}. The SR online algorithm generates a very accurate estimate of Spearman's
rank correlation even when the number of cutpoints is small ($m_1=m_2=30$). The
KT online algorithm seems to generate a very accurate estimate of the Kendall's
tau correlation when the absolute value of the correlation is small, but it seems to generate
a more biased estimate when the absolute value of the correlation is large.
This is because when $x$, $y$ are highly correlated, the $(x_i,y_i)$ pairs
are likely to be concentrated on the diagonal of matrix $M$. This concentration leads to a
poor approximation of the bivariate distribution of $(x,y)$ with matrix $M$,
which leads to biased estimates of Kendall's tau correlation. For the KT online
algorithm we suggest keeping the number of cutpoints above 100 in order to achieve
a more accurate result.

\section{Application to Sensor Data Generated in Industrial plant}
This section uses the proposed online algorithms to compute nonparametric correlations
based on sensor data that were generated in industrial plant from 2015 Prognostics and Health Management Society Competition \cite{phm}. The data contains sensor readings of 50 plants. For each plant, it provides sensor readings of four standard sensors S1-S4, and four control sensors R1-R4. We use the sensor readings of the first component in the first plant to do our experiment, where we compute nonparametric correlations based on sliding windows with window size 35,040 (corresponds to a one-year window).

\begin{figure}[ht]
	\centering
	\subfloat[Spearman's rank correlation]{
		\includegraphics[width=0.8\textwidth]{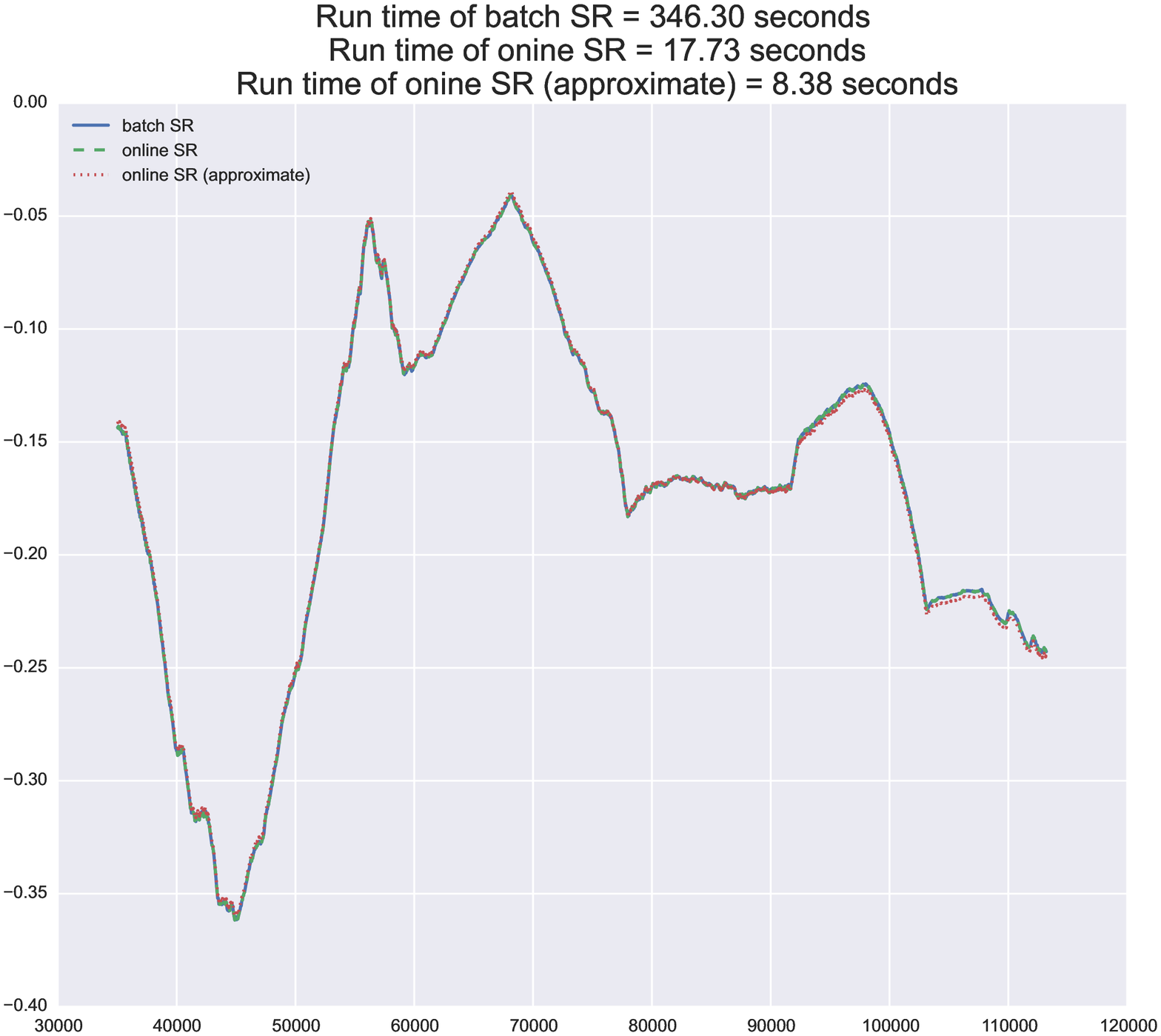}
		\label{fig:R1vsS1_SR}}
	
	\subfloat[Kendall's tau correlation]{
		\includegraphics[width=0.8\textwidth]{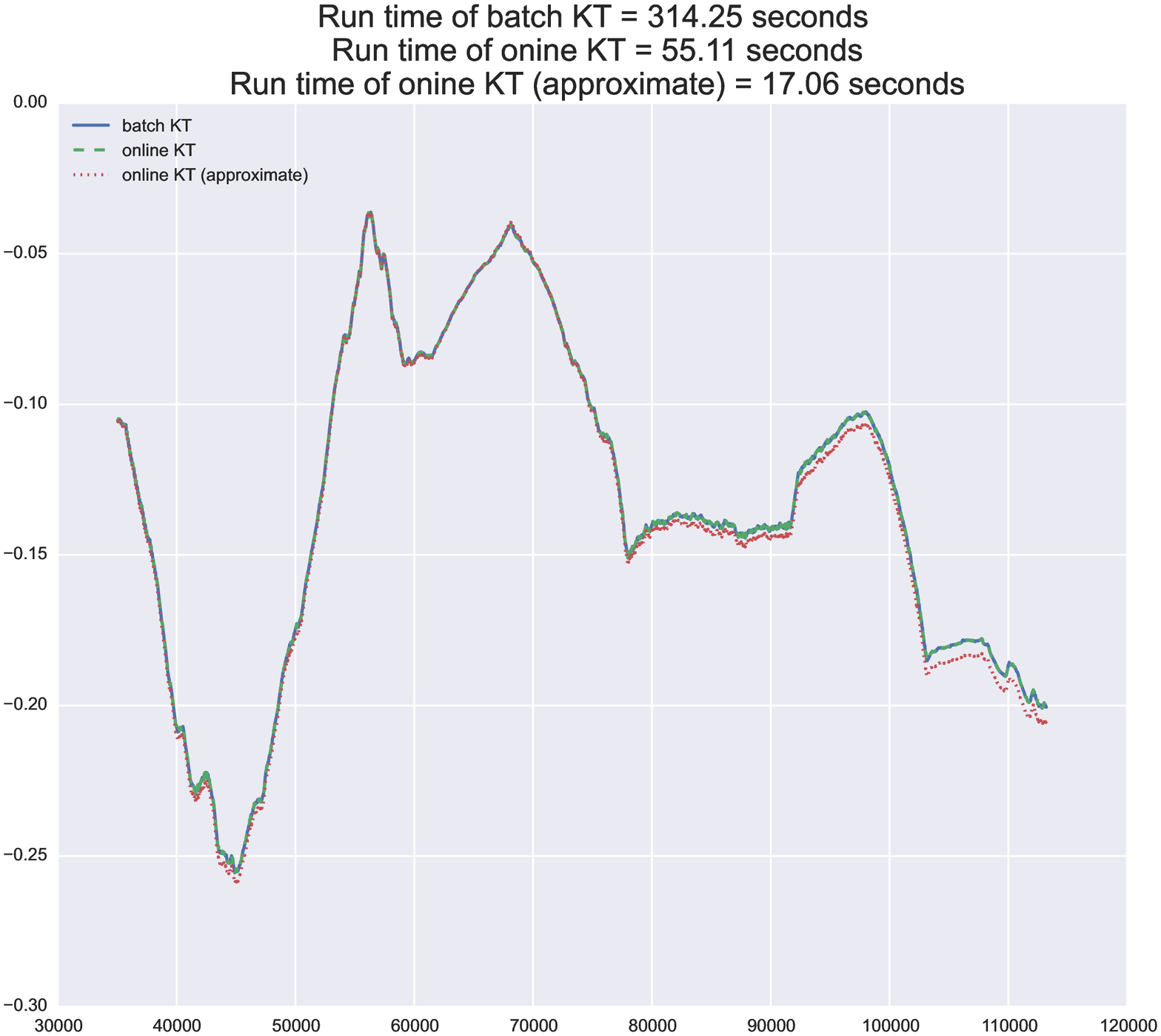}
		\label{fig:R1vsS1_KT}}
	\caption{Computation of nonparametric correlations between $R_1$ and $R_1$ with fixed size sliding windows.}
	\label{fig:R1vsS1}
\end{figure}

First, we compute the nonparametric correlation between $R_1$ and $S_1$. $R_1$ contains 10 unique values and we choose 9 cutpoints so that each unique value has its own cell in matrix $M$.  $S_1$ has 121 unique values, and we experiment on two methods to choose cutpoints for $S_1$. In the first method, we choose 120 cutpoints for $S_1$ so that each unique value of $S_1$ has its own cell in matrix $M$. In the second method, the cutpoints are chosen by first computing sample quantiles of $S_1$ at probabilities $0.05, 0.10, \ldots, 0.95$, and keeping only the
unique values. This leads to choosing 19 cutpoints for $S_1$. The result is shown in Figure~\ref{fig:R1vsS1}. We refer the result of the online algorithm with 19 cutpoints for $S_1$ as online SR (approximate) and online KT (approximate), respectively. Because the returned nonparametric correlations will only approximately equal the true values. We see a 20-50 fold speed up for SR and 5-20 fold speed up for KT.

\begin{figure}[ht]
	\centering
	\subfloat[Spearman's rank correlation]{
		\includegraphics[width=0.8\textwidth]{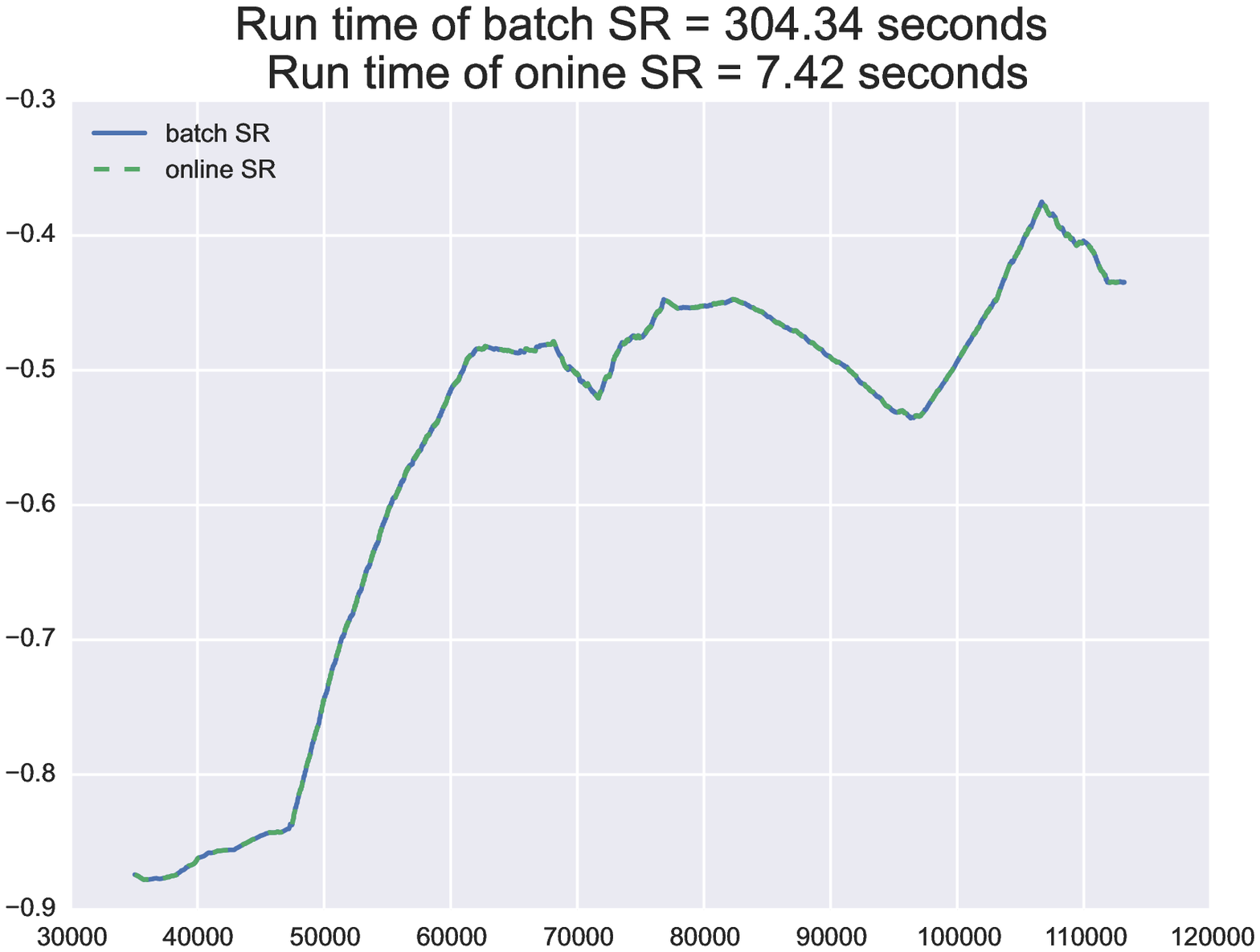}
		\label{fig:R3vsS3_SR}}
	
	\subfloat[Kendall's tau correlation]{
		\includegraphics[width=0.8\textwidth]{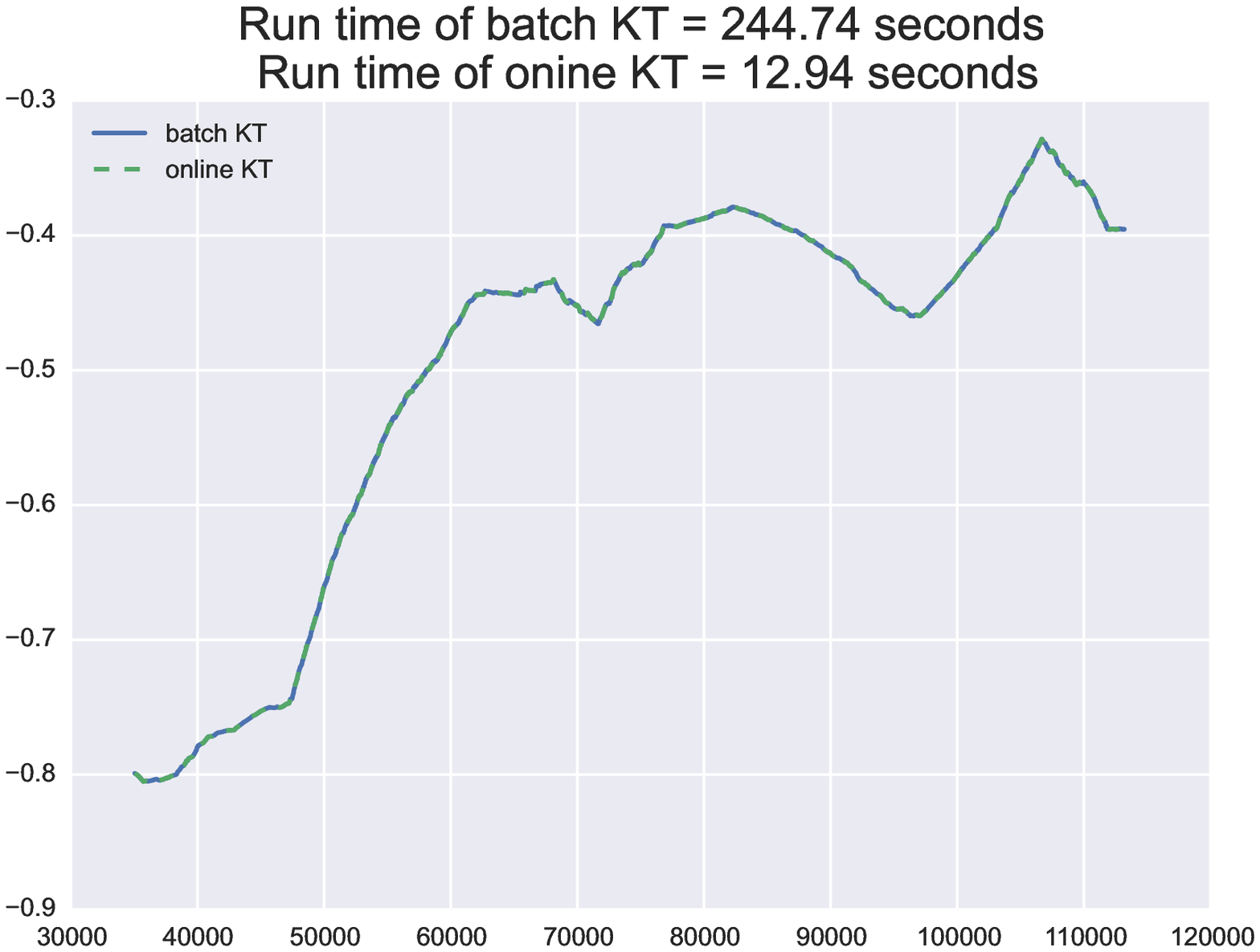}
		\label{fig:R3vsS3_KT}}
	\caption{Computation of nonparametric correlations between $R_2$ and $R_3$ with fixed size sliding windows.}
	\label{fig:R3vsS3}
\end{figure}

Then, we compute the nonparametric correlation between $R_3$ and $S_3$. We choose 7 cutpoints for $R_3$ and 11 cutpoints for $S3$ so that each unique value of $R_3$ and $S_3$ has its own cell in matrix $M$. The result is shown in Figure~\ref{fig:R3vsS3}. The online  algorithm returns the same result as that of batch algorithm with a 20-40 fold speed up.

\section{Conclusion}

This paper proposes a novel online algorithm for the computation of Spearman's
rank correlation and Kendall's tau correlation. The algorithm has time
complexity $O(1)$ and memory cost $O(1)$, and it is quite suitable for edge
devices, where only limited memory and processing power are available. By
changing the number of cutpoints specified in the algorithm, users can seek
a balance between speed and accuracy. The new online algorithm is very
fast and can easily compute the correlations 10 to 1,000 times faster than the corresponding batch algorithm (the number
varies over the settings of the problem). The
online algorithm can compute nonparametric correlations based either on all past
observations or on fixed-size sliding windows.

\bibliography{npcorr}
\bibliographystyle{ieeetr}

\end{document}